\documentclass[english,reprint, aps]{revtex4-1}
\usepackage{epsfig,amsmath}
\usepackage{subfigure}
\usepackage{graphicx}
\usepackage{dcolumn}
\usepackage{stmaryrd}
\usepackage{mathrsfs}
\usepackage{pifont}
\usepackage{amsthm}
\usepackage{amssymb}
\usepackage{bm}
\usepackage{latexsym}
\usepackage{color}
\usepackage{hyperref}
\usepackage{multirow}

\usepackage[english]{babel}
\usepackage{float}
\usepackage{hyperref}
\makeatother
\usepackage{color}
\usepackage{babel}
\usepackage {lineno}
 
\usepackage {footmisc}
\begin{document}

\renewcommand{\figurename}{Fig.}
\title{Control cost and quantum speed limit time in controlled almost exact state transmission in open systems}
\author{ Shen-Shuang Nie$^{1}$, Feng-Hua Ren$^{2}$, Run-Hong He$^{1}$, Jing Wu$^{1}$, }
\author{Zhao-Ming Wang$^{1}$}
\altaffiliation {Corresponding author: wangzhaoming@ouc.edu.cn}
\address {$^{1}$\mbox{College of Physics and Optoelectronic Engineering, Ocean University of China, Qingdao 266100, China }, $^{2}$\mbox{School of Information and Control Engineering, Qingdao University of Technology, Qingdao 266520, China}
}

\begin{abstract}
	
	We investigate the influence of environment noise on the control cost and the quantum speed limit time (QSLT) in the process of almost exact state transmission (AEST) through a spin chain under pulse control. The chain is immersed in its surrounding non-Markovian, finite temperature heat baths. We find that AEST can be realized in weak system-bath coupling, low temperature, and strong non-Markovian baths under effective external control. Correspondingly, the control cost and QSLT increases with increasing bath temperature and coupling strength. It is noticeable that non-Markovianity from the baths can be helpful to reduce the control cost and shorten the QSLT. Furthermore, we find that there exists a trade-off between the control cost and transmission fidelity and higher fidelity requires higher cost. In addition, the minimum control cost has been found to obtain certain transmission fidelity. 
	\begin{description}
		\item [{PACS~numbers}]  
	\end{description}

\end{abstract}
\maketitle

\section{INTRODUCTION}

Quantum state transfer (QST) has been widely investigated for its fundamental role to shuttle information between two quantum elements. For short distance communication, the spin chain as a communication channel has been extensively investigated since the Bose's work \cite{Bose2003}. However, normally the transmission fidelity decreases dramatically with increasing length of the chain \cite{Hu2009}. A lot of work has been done to improve the transmission fidelity through the chain. For example, the perfect state transfer (PST) can be obtained by properly engineering the coupling configuration of the chain \cite{Christandl2004,Christandl2005,Kay2006,Zhang2018}. High-fidelity QST in spin chain can also be achieved by applying appropriate external fields \cite{Fitzsimons2006,Balachandran2008}, or by applying an effective quantum control technique \cite{Wang2020a}. Experimentally, high-fidelity QST based on the Floquet-engineered method in a complex many-body system has been demonstrated \cite{Zhou2019}. In addition, conditional state transfer between quantum spin transistors in a Heisenberg spin chain has been realized \cite{Marchukov2016}.

In an actual system, the communication channel will have interactions with its surrounding environment, then the decoherence occurs and as a result the transmission fidelity decreases \cite{Jeske2013,Chen2016}. Quantum open systems have been used to describe the quantum systems with it immediate surroundings \cite{Li2019}. For a Markovian environment, where the environmental memory effects can be safely disregarded, Lindblad equations can be used to describe the system dynamics \cite{HeinzPeterBreuer2007,Hu2010}. When the memory effects cannot be neglected, a non-Markovian description of the system dynamics is required \cite{Ren2019}. The non-Markovianity of the environment will bring new physics to the open system \cite{Breuer2016,Vega2017,Roy2020}. For example, the memory effect of the non-Markovian environment can be used to generate macroscopic entanglement between two mirrors in optomechanical system \cite{ZhaoOE}. Non-Markovianity of the environment enhances the quality of the quantum correlation teleportation process with a couple of two-level atoms embedded in a zero-temperature bosonic bath \cite{Motavallibashi}. Measure for the degree of non-Markovianity is undoubtedly important, and different definitions are proposed \cite{Breuer2009,Chruscinski2014,Sampaio2019, Mirkin2019}. One of the widely accepted measure is based on the decreasing monotonicity of the trace distance under the completely positive and trace-preserving operations \cite{Breuer2009}. In a Markovian process, the monotonicity indicates that the information of distinguishability always flows from the system to environment, while for non-Markovian process, the monotonicity is violated and the information of distinguishability may flow back from the environment to the system \cite{Breuer2009}. Normally it is a difficult task to solve the non-Markovian dynamics. The quantum state diffusion (QSD) equation has been proved to be a promising approach to tackle this difficulty \cite{Ren2019, Shi2013,Yu1999,Li2020,Nakajima2018}. Analytical or exact solutions for many interesting systems have been obtained \cite{Diosi1998,Strunz1999}. Recently, by using the QSD approach, QST through a spin chain between two zero-temperature non-Markovian baths \cite{Ren2019} or in finite-temperature non-Markovian heat baths \cite{Wang2021} have been investigated. The transmission fidelity is found to decrease with increasing system-bath coupling strength, bath temperature, and bath Markovianity.

To combat the detrimental effects of the environment or the dispersion of the chain itself in the state transmission, an effective pulse control method has been proposed to realize almost exact state transmission (AEST) in a spin chain \cite{Wang2020,Wang2020a}. The pulses are realized by a leakage elimination operator (LEO) Hamiltonian and they must satisfy some certain conditions to guarantee the AEST \cite{Wang2020}. However, ``There ain't no such thing as a free lunch''. According to the second law of thermodynamics, non-ideal processes are always accompanied by irreversible consumption of thermodynamic resources \cite{Campbell2017,Landauer1961}. To obtain AEST via pulse control, the control cost must be considered. Studies on the fundamental lower bound \cite{Sagawa2008,Maroney2009,Piechocinska2000} of the thermodynamic energy cost provide a significant contribution to advance the molecular devices \cite{Joachim2000} and nanomachines \cite{Serreli2007}, since thermal and quantum fluctuations play considerable roles on the nanometer scale \cite{Zheng2016,Xiao2014}. Another fundamental bound is the maximal evolution speed of the quantum system \cite{Deffner2013}. The upper bound of state evolution speed, i.e., the quantum speed limit is established from the uncertain relationship between Heisenberg time and energy \cite{Pfeifer1993,Margolus1998,Giovannetti2004}. Quantum speed limit time (QSLT) represents the intrinsic minimal time interval for a quantum system evolving from an initial state to a target state. It has application in virtually all areas of quantum physics, such as quantum communication \cite{Bekenstein1981}, quantum computation \cite{Lloyd2000} and optimal control theory \cite{Caneva2009,Pyshkin2019}. QSLT in open systems has also been investigated \cite{Zhang2014,Campo2013,Taddei2013,Deffner2013,Deffner2013a}. In particular, a QSLT based on $p$ norm for an arbitrarily driven open system has been studied \cite{Deffner2013}. The memory effect of environment characterized by non-Markovianity has been found to shorten QSLT in a single qubit \cite{Deffner2013} or multiqubit \cite{Zhu2015} case.

 For open systems, the bath is usually assumed to be collective for simplicity. However, under realistic conditions such as in adiabatic quantum computation \cite{Zhao2019,Albash2015} 
, it is more practical to assume that each qubit interacts with its individual bath. In this paper, we first investigate the AEST through a spin chain by external pulse control. Each spin encounters its individual non-Markovian and finite temperature heat bath. Using the strategy in Ref.~\cite{Wang2020}, adding an LEO Hamiltonian to the system, we find that AEST can be obtained in finite temperature heat baths with weak system-bath coupling, strong non-Markovianity and low temperature. Then we study the influence of the environment noise on the control cost and QSLT for the achievement of AEST. We find that non-Markovianity of the baths plays an essential role both for reducing the control cost and shortening the QSLT.   

\section{MODEL AND HAMILTONIAN}
Considering a quantum system embedded in $N$-independent multimode bosonic baths, the total Hamiltonian can be written as
\begin{equation}
H_{tot}=H_{s}+H_{b}+H_{int},
\end{equation}
where $H_{s}$ is the system Hamiltonian and $H_{b}=\sum_{j=1}^{N}H_{b}^{j}$ is the $N$-independent baths Hamiltonian with the $j$th bath's Hamiltonian $H_{b}^{j}=\sum_{k}\omega_{k}^{j}b_{k}^{j\dagger}b_{k}^{j}$ (setting $\hbar=1$ from now on). $b_{k}^{j\dagger}$, $b_{k}^{j}$ represent the bosonic creation and annihilation operators and $\omega_{k}^{j}$ refers to the bosonic frequency of the $j$th bath. The interaction Hamiltonian is
\begin{equation}
H_{int}=\sum_{j,k}(g_{k}^{j*}L_{j}^{\dagger}b_{k}^{j}+g_{k}^{j}L_{j}b_{k}^{j\dagger}),
\end{equation}
where $L_{j}$ is the Lindblad operator. It describes the coupling between the system and the $j$th bath. $g_{k}^{j}$ is the coupling constant between the system and $k$th model of $j$th bath.

Suppose that the initial state of the $j$th bath is in a thermal equilibrium state at temperature $T_{j}$ and the density operator reads 
 
\begin{equation}
\rho_{j}\left(0\right)=e^{-\beta_{j} H_{b}^{j}}/Z_{j},
\end{equation}
where $Z_{j}=\mathrm{Tr}[e^{-\beta_{j} H_{b}^{j}}]$ is the partition function and $\beta_{j}=1/K_{B}T_{j}$. The system dynamics can be calculated by the QSD approach. The total wave function $|\Psi_{tot}\left(t\right)\rangle$ is projected to the coherent state of the bath modes $|z\rangle$ and $|w\rangle$, $|\Psi_{t}\left(z_{1}^{*},w_{1}^{*},\ldots,z_{N}^{*},w_{N}^{*}\right)\rangle=\langle z_{1},w_{1},\ldots z_{N},w_{N}|\Psi_{tot}\left(t\right)\rangle$, which is known as the stochastic quantum trajectory. It obeys a linear QSD equation \cite{Diosi1998}
\begin{equation}
\frac{\partial|\Psi_{t}\rangle}{\partial t}= [-iH_{s}+\underset{j}{\sum}(L_{j}z_{t}^{j*}-L_{j}^{\dagger}\overline{O}_{z}^{j}+L_{j}^{\dagger}w_{t}^{j*}-L_{j}\overline{O}_{w}^{j})]|\Psi_{t}\rangle, \label{eq:4} 
\end{equation}
where
\begin{equation}
 z_{t}^{j*}=-i\sum_{k}\sqrt{\overline{n}_{k}^{j}+1}g_{k}^{j}z_{k}^{j*}e^{i\omega_{k}^{j}t},
\end{equation}
\begin{equation}
w_{t}^{j*}=-i\sqrt{\overline{n}_{k}^{j}}g_{k}^{j*}w_{k}^{j*}e^{i\omega_{k}^{j}t},	
\end{equation}
are the $j$th independent complex Gaussian processes. $\overline{n}_{k}^{j}=\frac{1}{\mathrm{exp}(\hbar\omega_{k}^{j}/K_{B}T_{j})}$ is the mean thermal occupation number of quanta in mode $\omega_{k}^{j}$. The operators ${O}_{z,(w)}^{j}$ are defined by the $ 
\frac{\delta\left|\Psi_{t}\right\rangle}{\delta z_{s}^{j},(w_{s}^{j})}=O_{z,(w)}^{j}\left(t, s, z_{1}^{*}, w_{1}^{*}, \ldots \right)\left|\Psi_{t}\right\rangle$ \cite{Diosi1998,Yu2004}, and $\overline{O}_{z,(w)}^{j}=\int_{0}^{t}ds\alpha_{z,(w)}^{j}(t-s)O_{z,(w)}^{j}$. Here $\alpha_{z,(w)}^{j}(t-s)$ are the correlation functions. 

From the consistency condition, the operators ${O}_{z,(w)}^{j}$ satisfy \cite{Yu2004}
\begin{eqnarray}
\frac{\partial O_{z}^{j}}{\partial t} &=&[-iH_{s}+\sum_{j}(L_{j}z_{t}^{j\ast }-L_{j}^{\dag }\overline{O}%
_{z}^{j}+L_{j}^{\dag }w_{t}^{j\ast }-L_{j}\overline{O}_{w}^{j}),\nonumber\\&\;&
O_{z}^{j}]-\sum_{j}(L_{j}^{\dag }\frac{\delta \overline{O}_{z}^{j}%
}{\delta z_{s}^{j\ast }}+L_{j}\frac{\delta \overline{O}_{w}^{j}}{\delta
	z_{s}^{j\ast }}),  \label{eq:7}
\end{eqnarray}

\begin{eqnarray}
\frac{\partial O_{w}^{j}}{\partial t} &=&[-iH_{s}+\sum_{j}(L_{j}z_{t}^{j\ast }-L_{j}^{\dag }\overline{O}%
_{z}^{j}+L_{j}^{\dag }w_{t}^{j\ast }-L_{j}\overline{O}_{w}^{j}),\nonumber\\&\;&
O_{w}^{j}]-\sum_{j}(L_{j}^{\dag }\frac{\delta \overline{O}_{z}^{j}%
}{\delta w_{s}^{j\ast }}+L_{j}\frac{\delta \overline{O}_{w}^{j}}{\delta
	w_{s}^{j\ast }}).  \label{eq:8}
\end{eqnarray}

Now we use Eqs.~\eqref{eq:7} and~\eqref{eq:8} to derive a non-Markovian master equation. We define $P_{t}\equiv|\Psi_{t}\rangle\langle\Psi_{t}|$ and the reduced density matrix of the system $\rho_{s}=M[P_{t}]$. Here $M[P_{t}]$ is given by the ensemble average over all of the quantum trajectories. According to Refs.~\cite{Yu2004,Ren2020}, the evolution equation of the system density matrix reads
\begin{eqnarray}
\frac{\partial}{\partial t}\rho_{s}&=&-i\left[H_{s},\rho_{s}\right]+\mathop{\underset{j}{\sum}}\{[L_{j},M[P_{t}\overline{O}_{z}^{j\dagger}]]-[L_{j},M[\overline{O}_{z}^{j}P_{t}]]\nonumber\\&\;&
 +[L_{j}^{\dagger},M[P_{t}\overline{O}_{w}^{j\dagger}]]-[L_{j},M[\overline{O}_{w}^{j}P_{t}]]\}.\label{eq:9}
\end{eqnarray}
 
 The above equation is the master equation in finite-temperature baths. However, this equation is not a closed equation for $\rho_{s}$.  The $\overline{O}_{z, (w)}^{j}$ operators generally contain noises except for some special cases. If the baths couple to the system weakly, we can assume $\overline{O}_{z, (w)}^{j}\left(t,z_{1}^{*},w_{1}^{*},\ldots,z_{N}^{*},w_{N}^{*}\right)=\overline{O}_{z, (w)}^{j}\left(t\right)$, then Eq.~\eqref{eq:9} becomes a closed master equation  \cite{Wang2021}
\begin{eqnarray}
\frac{\partial}{\partial t}\rho_{s}&=& -i[H_{s},\rho_{s}]+\underset{j}{\sum}\{[L_{j},\rho_{s}\overline{O}_{z}^{j\dagger}\left(t\right)]-[L_{j}^{\dagger},\overline{O}_{z}^{j}\left(t\right)\rho_{s}]\vspace{1ex}\nonumber\\&\;&
+[L_{j}^{\dagger},\rho_{s}\overline{O}_{w}^{j\dagger}\left(t\right)]-[L_{j},\overline{O}_{w}^{j}\left(t\right)\rho_{s}]\}.
\label{eq:10}
\end{eqnarray}

 In order to calculate the correlation function, we need to introduce the spectral density. Here we use spectrum density $J(\omega_{j})=\frac{\Gamma_{j}}{\pi}\frac{\omega_{j}}{1+(\omega_{j}/\gamma_{j})^{2}}$, which is Ohmic type with a Lorentz-Drude cutoff \cite{Meier}, where $\Gamma_{j}$, $\gamma_{j}$ are real parameters. $\Gamma_{j}$ represents the coupling strength between the system and $j$th bath and $\gamma_{j}$ is the characteristic frequency of the $j$th bath. For weak system-bath coupling, $\Gamma_{j}\ll1$, we have \cite{Wang2021} 
\begin{equation}
\alpha_{z}^{j}\left(t-s\right)=\Gamma_{j}T_{j}\Lambda_{j}\left(t,s\right)+i\Gamma_{j}\dot{\Lambda}_{j}\left(t,s\right)\label{eq:11},
\end{equation}
\begin{equation}
\alpha_{w}^{j}\left(t-s\right)=\Gamma_{j}T_{j}\Lambda_{j}\left(t,s\right)\label{eq:12},
\end{equation}
where $\Lambda_{j}\left(t,s\right)=\frac{\gamma_{j}}{2}e^{-\gamma_{j}|t-s|}$ is an Ornstein-Uhlenbeck correlation function and it decays on the environmental memory time $1/\gamma_{j}$. For $\gamma_{j}\rightarrow\infty$, it corresponds to a Markovian limit. Now the two correlation functions in Eqs. \eqref{eq:11} and \eqref{eq:12} satisfy
\begin{equation}
\frac{\partial\alpha_{z\left(w\right)}^{j}}{\partial t}=-\gamma_{j}\alpha_{z\left(w\right)}^{j}\left(t-s\right).
\end{equation}

From the above relations, the $\overline{O}_{z,(w)}^{j}$ operators satisfy the following equations \cite{Wang2021}

\begin{eqnarray}
\frac{\partial\overline{O}_{z}^{j}}{\partial t}&=& (\frac{\Gamma_{j}T_{j}\gamma_{j}}{2}-\frac{i\Gamma_{j}\gamma_{j}^{2}}{2})L_{j}-\gamma_{j}\overline{O}_{z}^{j}\vspace{1ex}\nonumber\\&\;&
  +[-iH_{s}-\underset{j}{\sum}(L_{j}^{\dagger}\overline{O}_{z}^{j}+L_{j}\overline{O}_{w}^{j}),\overline{O}_{z}^{j}],
  \label{eq:14}
\end{eqnarray}
\begin{eqnarray}
\frac{\partial\overline{O}_{w}^{j}}{\partial t}&=& \frac{\Gamma_{j}T_{j}\gamma_{j}}{2}L_{j}^{\dagger}-\gamma_{j}\overline{O}_{w}^{j}\nonumber\\&\;&
 +[-iH_{s}-\underset{j}{\sum}(L_{j}^{\dagger}\overline{O}_{z}^{j}+L_{j}\overline{O}_{w}^{j}),\overline{O}_{w}^{j}].
 \label{eq:15}
\end{eqnarray}
The master equation Eq.~\eqref{eq:10} can be numerically calculated by using Eqs. \eqref{eq:14} and \eqref{eq:15}. In the Markovian limit, the system density matrix reduces to the Lindblad equation \cite{Wang2021} 
\begin{eqnarray}
\frac{\partial}{\partial t}\rho_{s}&=&-i[H_{s},\rho_{s}]\nonumber\\&\;&+\sum_{j}\frac{\Gamma_{j}T_{j}}{2}[(2L_{j}\rho_{s}L_{j}^{\dagger}-L_{j}^{\dagger}L_{j}\rho_{s}-\rho_{s}L_{j}^{\dagger}L_{j})\nonumber\\&\;&+(2L_{j}^{\dagger}\rho_{s}L_{j}-L_{j}L_{j}^{\dagger}\rho_{s}-\rho_{s}L_{j}L_{j}^{\dagger})].\label{eq:16}
\end{eqnarray}

Here we consider a one-dimensional XY spin chain as the communication channel 
\begin{equation}
H_{s}=\sum_{i=1}^{N-1}J_{i,i+1}\left(\sigma_{i}^{x}\sigma_{i+1}^{x}+\sigma_{i}^{y}\sigma_{i+1}^{y}\right)\label{eq:17},
\end{equation}
where $J_{i,i+1}$ is the coupling strength between the nearest neighbour sites $i$ and $i+1$. $\sigma_{i}^{k} (k=x,y)$ are the Pauli operators acting on the $i$th spin. We set $J_{i,i+1}=-1$, and it corresponds to a uniform chain with open boundary conditions.

 Suppose initially the system is prepared at the state $|\Psi_{s}\left(0\right)\rangle=|1\cdots00\rangle$, and the target task is to transfer the state $|1\rangle$ to the other end of the chain at time $t$ with $|\Psi_{s}\left(t\right)\rangle= |0\cdots01\rangle$. Here the size of the system is typically small so that each sites are close to share the same environment parameters. Therefore we assume that the parameters of $N$ independent baths are taken as same, $\Gamma_{i}=\Gamma$, $\gamma_{i}=\gamma$, $T_{i}=T$ for $i=1,2,3\ldots N$ \cite{Wang2021}. We emphasize that even all the parameters are assumed to be equal, it is different from the one single bath with $\Gamma, \gamma$ and $T$ case. For $N$-independent baths model, there are $N$ coupled equations in total for $\overline{O}_{z,(w)}^{j}$ operators. As a comparison, there would be just one equation for a single common bath. The fidelity at time $t$ can be defined as $F\left(t\right)=\sqrt{\langle\Psi_{s}\left(t\right)|\rho_{s}\left(t\right)|\Psi_{s}\left(t\right)\rangle}$, where $\rho_{s}\left(t\right)$ is the system density matrix in Eq.~$\eqref{eq:10}$. Now the initial state of the whole system is
\begin{equation}
\rho_{tot}\left(0\right)=\rho_{s}\left(0\right)\otimes\rho_{1}\left(0\right)\otimes\cdots\otimes\rho_{N}\left(0\right),
\end{equation}
where $\rho_{j}\left(0\right)$ indicates that the $j$th bath is in a thermal equilibrium state at time $t=0$.

\section{AEST in finite temperature heat baths under pulse control}

Dispersion effects and the environmental noise often degrade the transmission fidelity \cite{Hu2009,Hu2010}. Pulse control technique has been used to realize the AEST \cite{Wang2020a,Wang2020}, adiabatic speedup \cite{Wang2018}, adiabatic QST \cite{Chen2014,Chen2018,Huang2018} and adiabatic quantum computation \cite{Wang2016,Ren2020}. In this part we investigate AEST through a uniform spin chain in finite temperature heat baths by pulse control. The basic idea is letting the system go along the AEST passage by adding an LEO Hamiltonian to the system’s dynamics. 

At first we construct an LEO Hamiltonian as in Ref.~\cite{Wang2020}. This will be achieved by considering PST Hamiltonian ($H_{PST}$), whose coupling strength $J_{i,i+1}=-\sqrt{i\left(N-i\right)}$ in Eq.~\eqref{eq:17}. Then the PST can be obtained, at time $t=n\pi/4$ ($n$ is an odd integer), by the PST couplings driving \cite{Christandl2004}. The initial state of the system is still $|\Psi_{s}\left(0\right)\rangle$ and its evolution is denoted as $|\psi_{1}\left(t\right)\rangle=\mathrm{exp}\left(-iH_{PST}t\right)|\Psi_{s}\left(0\right)\rangle$.  The LEO Hamiltonian can be constructed as \cite{Wang2020}
\begin{equation}
H_{LEO}=c\left(t\right)|\psi_{1}\left(t\right)\rangle\langle\psi_{1}\left(t\right)|,
\end{equation}
where $c\left(t\right)$ is the control function. The LEO Hamiltonian can be realized by a sequence of pulses and how to apply the control has been discussed in Ref.~\cite{Wang2020}. Then the total Hamiltonian is
\begin{equation}
H=H_{tot}+H_{LEO}.
\end{equation}

In this paper, we consider nonperturbative control pulse which is tunable and finite both on the strength and duration. We choose the zero-area pulse sequence \cite{Chen2018,Zhang2019} and the pulse takes alternating positive and negative values. The characteristic of such sequence is its net area is always zero in a control period. For the rectangular pulses, the control function can be taken as \cite{Pyshkin2016}
\begin{equation}
c(t)=\left\{ \begin{array}{ll}
I\begin{array}{c}
,\end{array}n\tau<t<(n+1)\tau, & n\text{ is even }\\
-I, & \text{ otherwise },
\end{array}\right.\label{eq:21}
\end{equation}
where $I$ and $\tau$ are the pulse intensity and half-period, respectively. The effective control can be obtained when the pulses satisfy $I\tau=2\pi m$ with $m=1,2,3,\cdots$ \cite{Pyshkin2016}. Similarly, for the sine function \cite{Wang2020b}, $c\left(t\right)=I\mathrm{sin}\left(\omega t\right)$ with $\omega \tau=\pi$, the effective control condition is $J_{0}\left(I\tau/\pi\right)=0$. $J_{0}\left(x\right)$ is the zero order Bessel function of the first kind \cite{Wang2018,Wang2020}. 
\begin{figure}
	\centerline{\includegraphics[width=1.0\columnwidth]{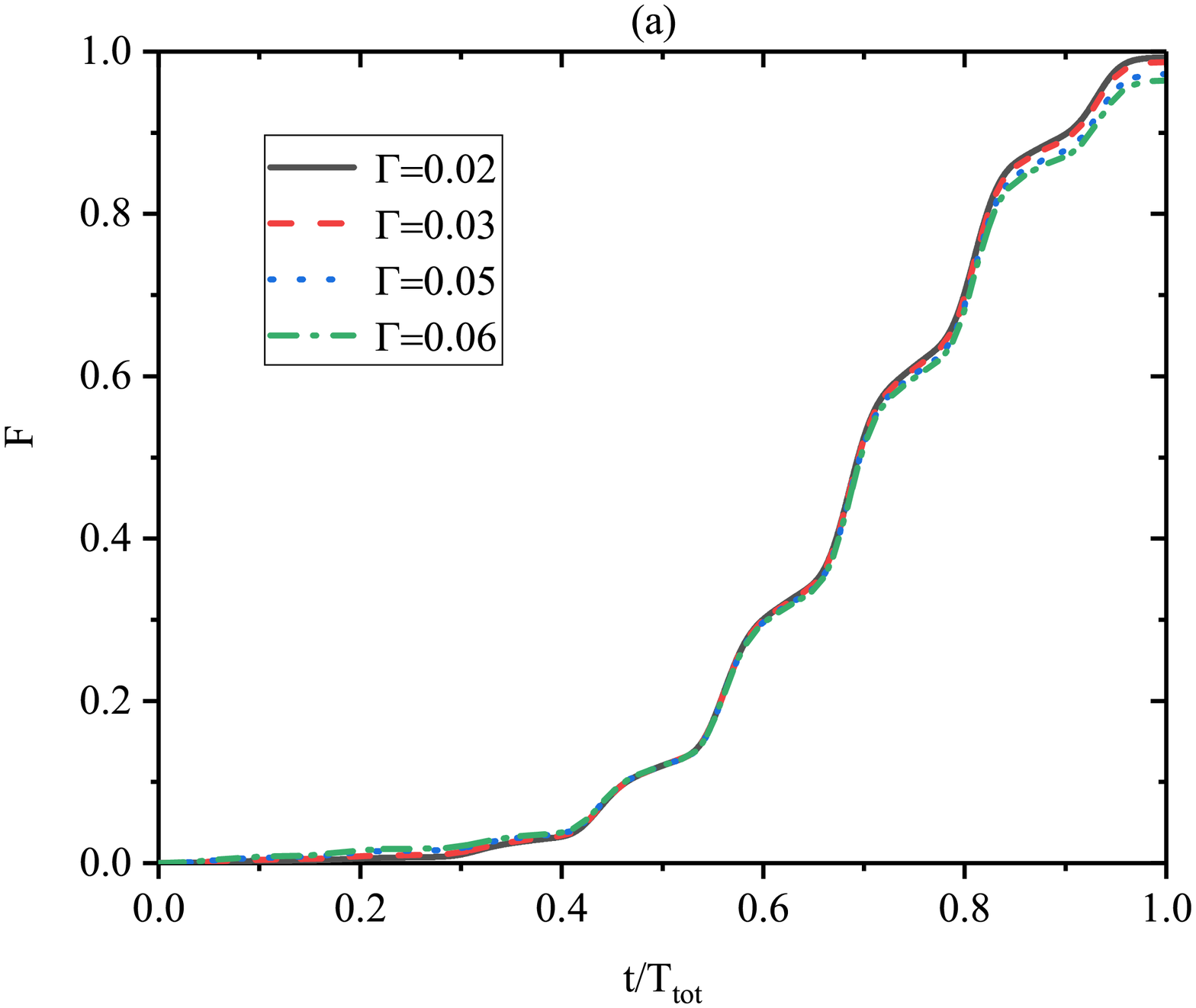}}
	\centerline{\includegraphics[width=1.0\columnwidth]{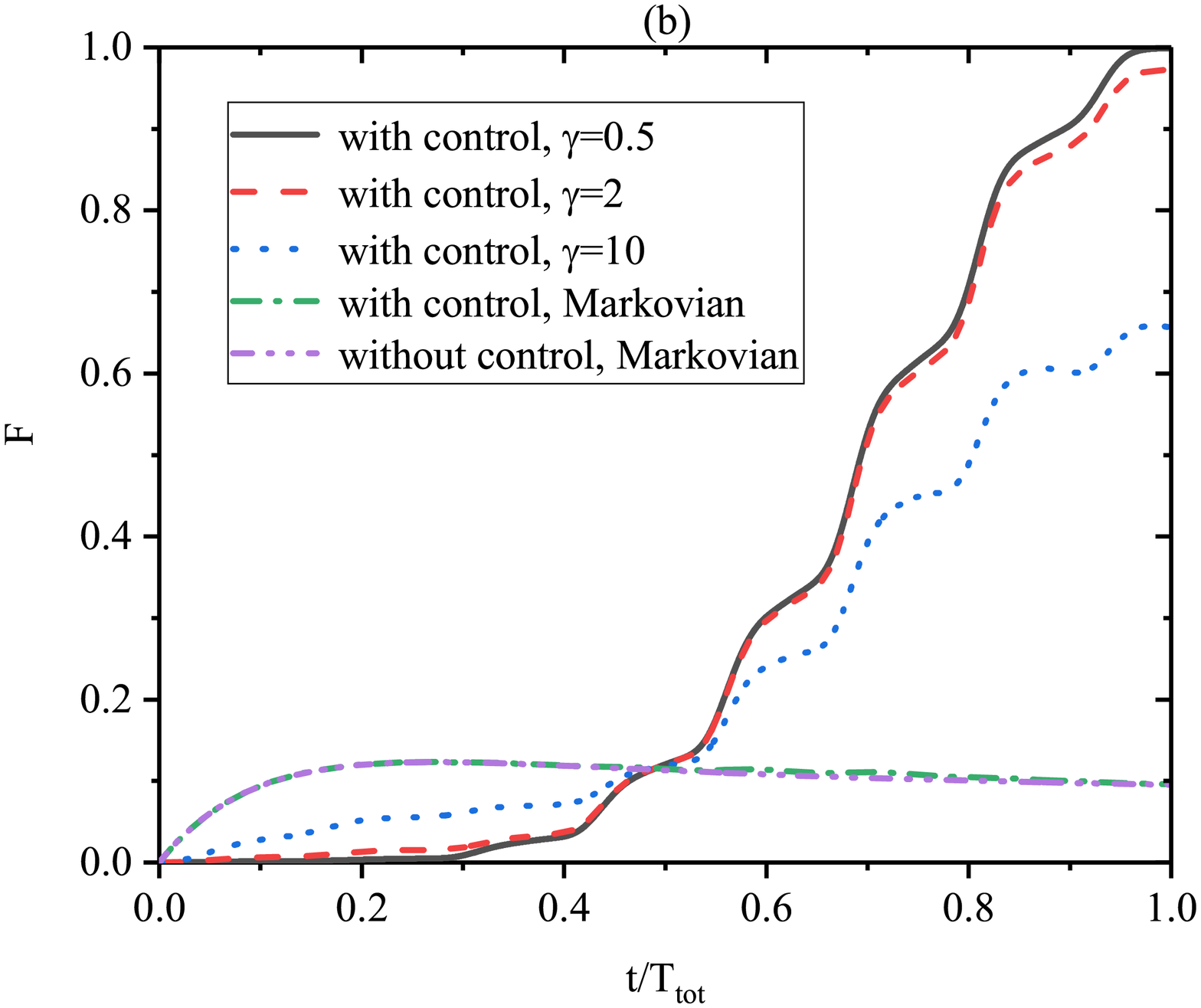}}
	\centerline{\includegraphics[width=1.0\columnwidth]{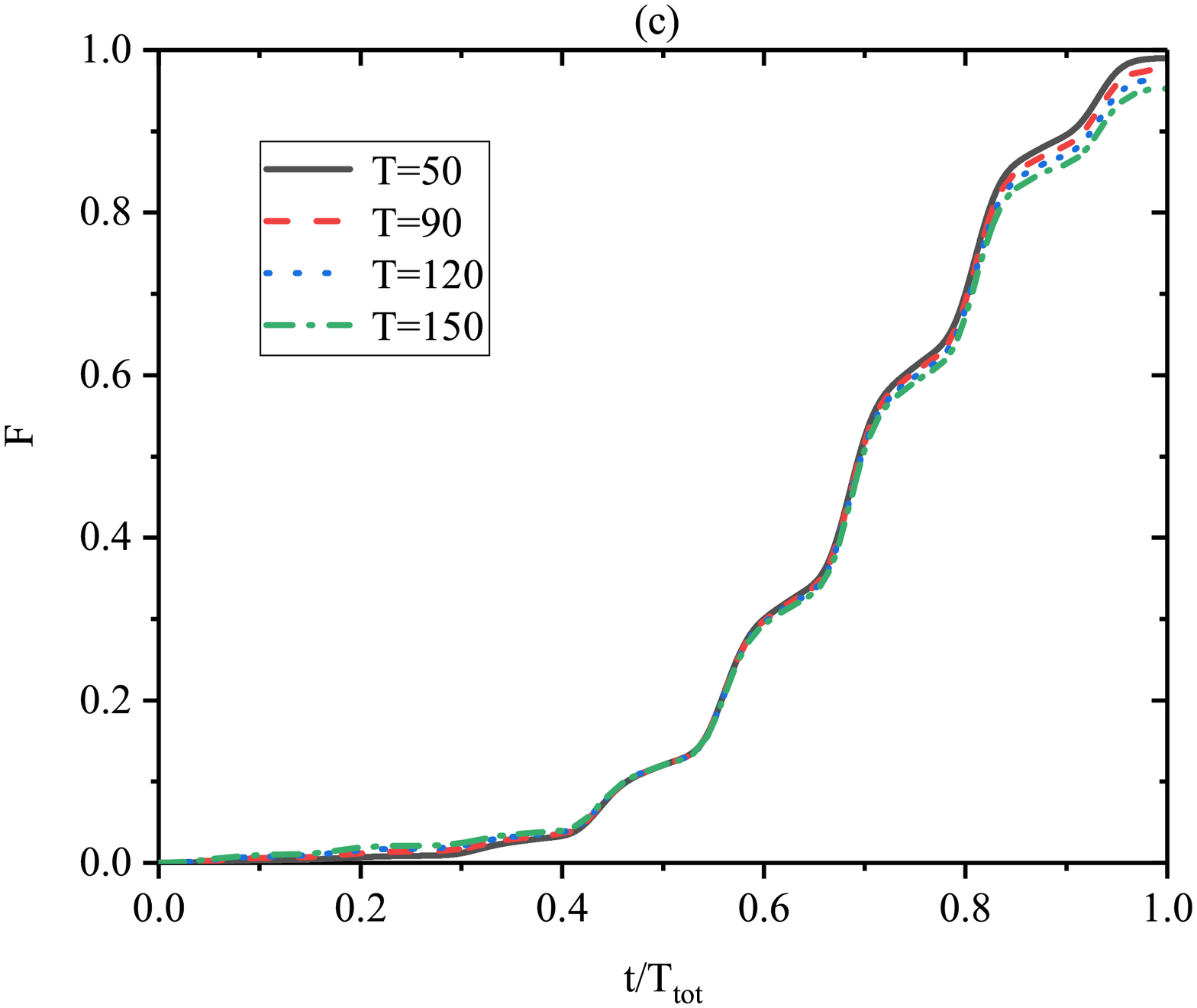}}
	\caption{(Color on line) Controlled AEST in $L=\sigma^{-}$ open system: the fidelity $F$ as a function of rescaled time $t/T_{tot}$ for different environmental parameters (a)  $\Gamma$, $\gamma=2$, $T=40$; (b) $\gamma$, $\Gamma=0.04$, $T=50$; (c) $T$, $\Gamma=0.02$, $\gamma=2$. The total evolution time is $T_{tot}=\pi/4$ for the controlled or free evolution. The control function is $c(t)=I sin(\omega t)$ with $\omega \tau =\pi$. $\tau=\pi/32$, $I=76.96$, which satisfies the condition $I\tau=2.405\pi$. The length of the chain is chosen to be $N=7$.}
	
	\label{fig:1}	
	
\end{figure} 

First we use the sine function as an example. The pulse parameters are taken as $\tau=\pi/32$ and $I=76.96$ which satisfy $I\tau=2.405\pi$. Note here 2.405 is the zero point of the zero order Bessel function. In Fig.~\ref{fig:1}, we plot the fidelity $F$ versus the rescaled time $t/T_{tot}$ for different parameters $\Gamma$, $\gamma$ and $T$, respectively. We take $L=\sigma^{-}=\sum_{j=1}^{N}\sigma_{j}^{-}$ as an example, where $\sigma_{j}^{-}$ denotes the lowering operator for the $j$th spin. The total evolution time is taken as $T_{tot}=\pi/4$ throughout the paper, which corresponds to a fixing actual driving time. Also PST is realized at this time when the chain has the PST coupling \cite{Christandl2005}. The length of the chain is chosen to be $N=7$. In Fig.~\ref{fig:1}(a) we plot the influence of the system-bath coupling strength $\Gamma$ to the transmission fidelity $F$. $\gamma=2$, $T=40$. It shows that when adding the pulse, AEST can be obtained. When $\Gamma=0.06$, $F\approx0.96$ at $t=\pi/4$. In Fig.~\ref{fig:1}(b) we consider different parameters $\gamma$. $\Gamma=0.04$, $T=50$. The results show that non-Markovianity from the baths plays an essential role in boosting the fidelity. AEST can only be obtained in strong non-Markovian baths. For example, $\gamma=0.5$, $F=0.999$. It is in accordance with Ref.~\cite{Wang2020a} that memory effects can boost the effectiveness of LEO control pulses in a zero-temperature bath. Clearly the control fails to boost the fidelity in Markovian baths. In Fig.~\ref{fig:1}(b) We also plot the comparison of the fidelity evolution with and without control in Markovian baths and find that the fidelity is almost equal. This can be explained by the fact that for a Markovian bath the correlation function becomes a $\delta$-function, then gives   $\overline{O}_{z}^{j}=\frac{\Gamma_{j}T_{j}}{2}L_{j}$,  $\overline{O}_{w}^{j}=\frac{\Gamma_{j}T_{j}}{2}L_{j}^{\dagger}$. The control Hamiltonian $H_{LEO}$ will fail to affect $\overline{O}_{z,(w)}^{j}$ operators, and as a result, the control loses its effectiveness \cite {Wang2020a}. Fig.~\ref{fig:1}(c) plots the time evolution of the fidelity for different temperature $T$. $\Gamma=0.02$, $\gamma=2$. As expected the fidelity is more higher in a lower temperature. AEST can even be realized for a higher temperature as long as the bath is non-Markovian. For example, $T=150$, $F(T_{tot})=0.952$.

\section{TRADE-OFF BETWEEN CONTROL COST AND TRANSMISSION FIDELITY}

When adding the external driving, the control cost must be considered. A family of cost function has been proposed \cite{Zheng2016} to quantify the driving cost. One of the cost function can be defined as $C_{t}^{n}\equiv\nu_{t,n}\int_{0}^{\tau}\mathrm{d}t\left\Vert H_{LEO}(t)\right\Vert $, where $\left\Vert A\right\Vert =\sqrt{\mathrm{tr}\left\{ A^{\dagger}A\right\} }$  is the  Hilbert-Schmidt norm \cite{Santos2015}, and in the following we set $\nu_{t,n}=n=1$ for simplicity \cite{Zheng2016}, 
\begin{equation}
C_{t}^{1}\equiv C=\int_{0}^{\tau}dt\left\Vert H_{LEO}(t)\right\Vert. 
\end{equation}
 The cost $C$ can be viewed as the additional effect of control pulse \cite{Zheng2016}. Therefore it is easy to see that the instantaneous cost of the control pulse is
\begin{equation}
\partial_{t}C=\left\Vert H_{LEO}(t)\right\Vert =\left|c\left(t\right)\right|\sqrt{\underset{n}{\sum}\left|\langle n|\psi_{1}\left(t\right)\rangle\right|^2}.\label{eq:23}
\end{equation}
\begin{figure}
	\centerline{\includegraphics[width=1.0\columnwidth]{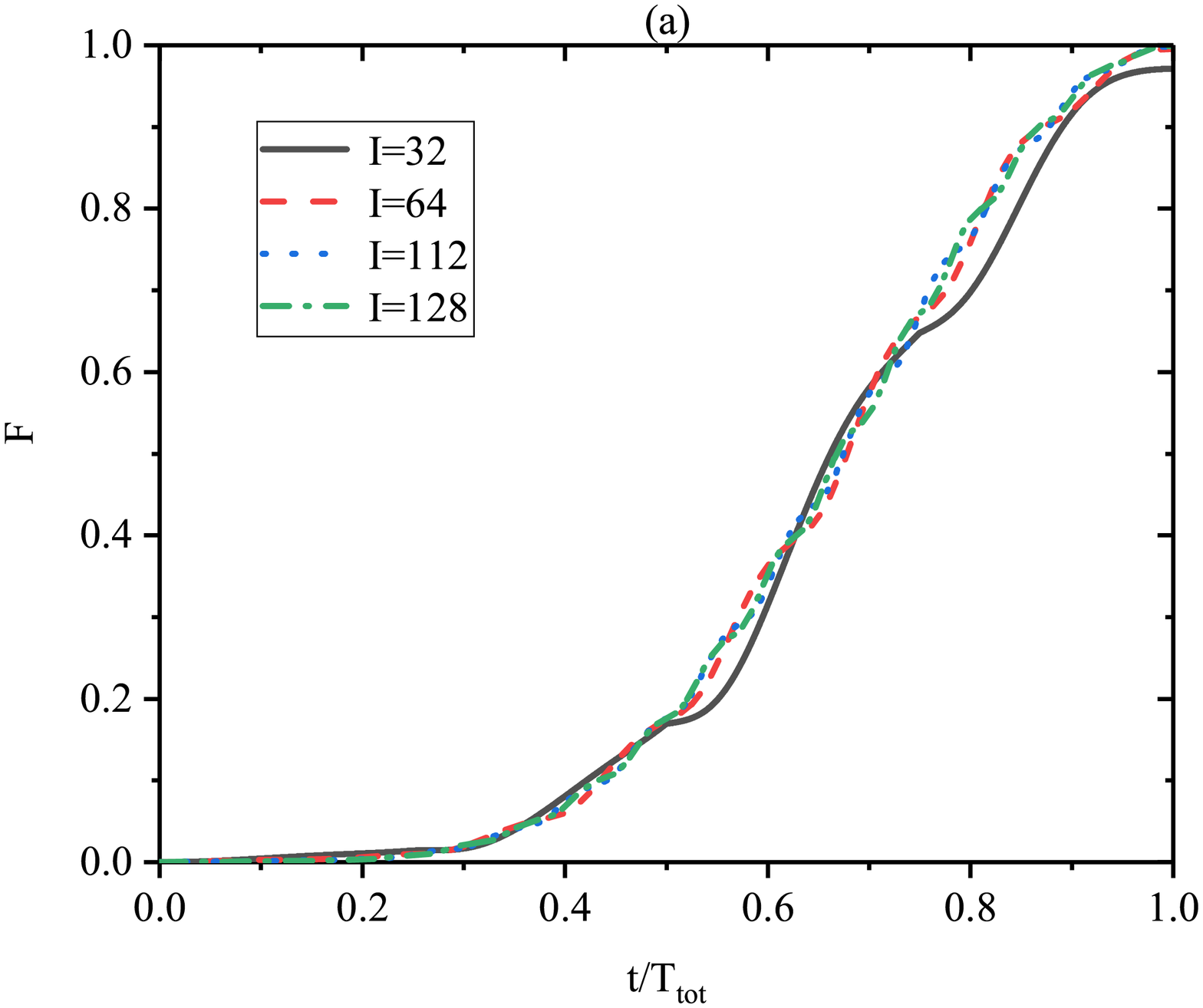}}
	\centerline{\includegraphics[width=1.0\columnwidth]{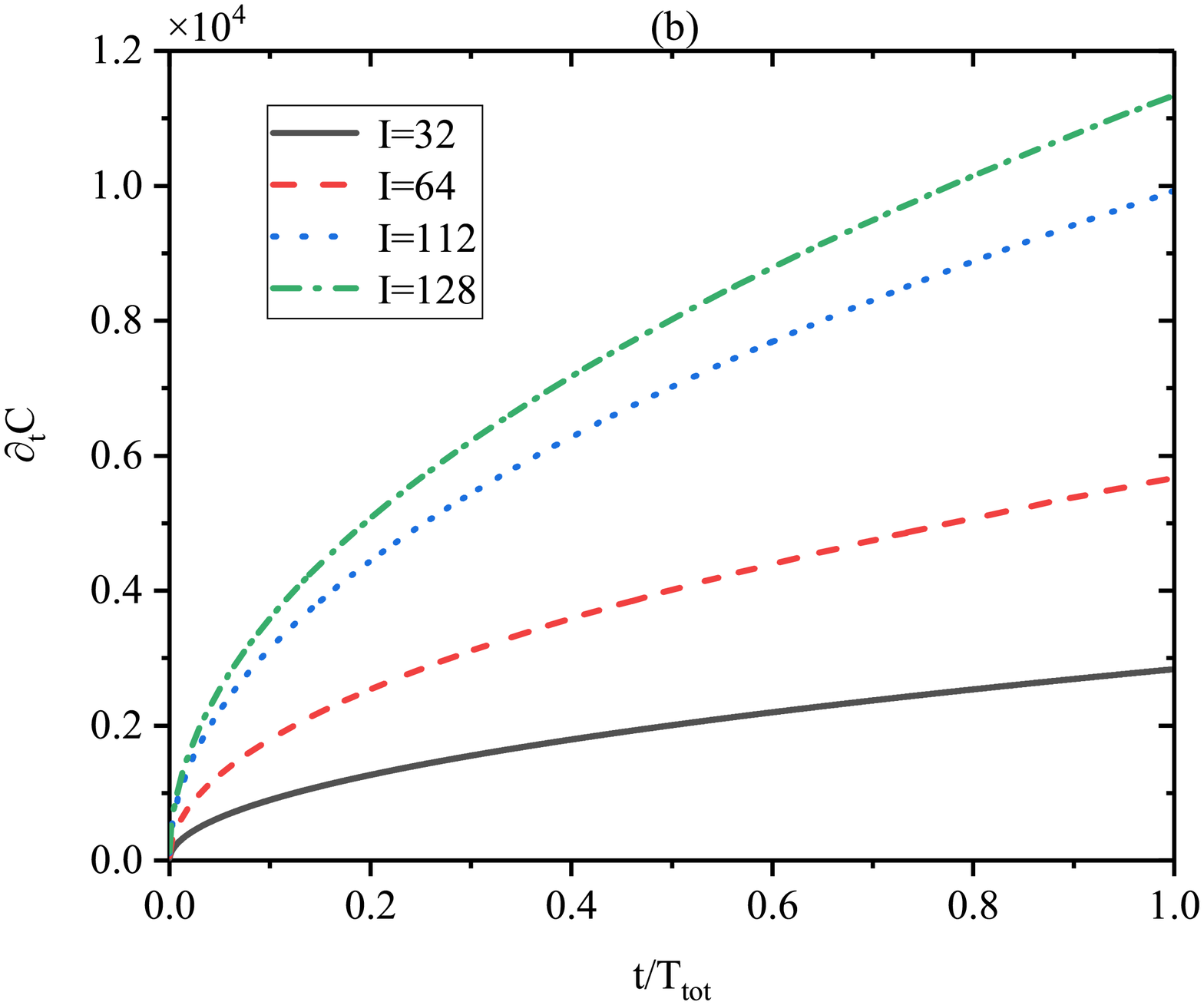}}
	\centerline{\includegraphics[width=1.0\columnwidth]{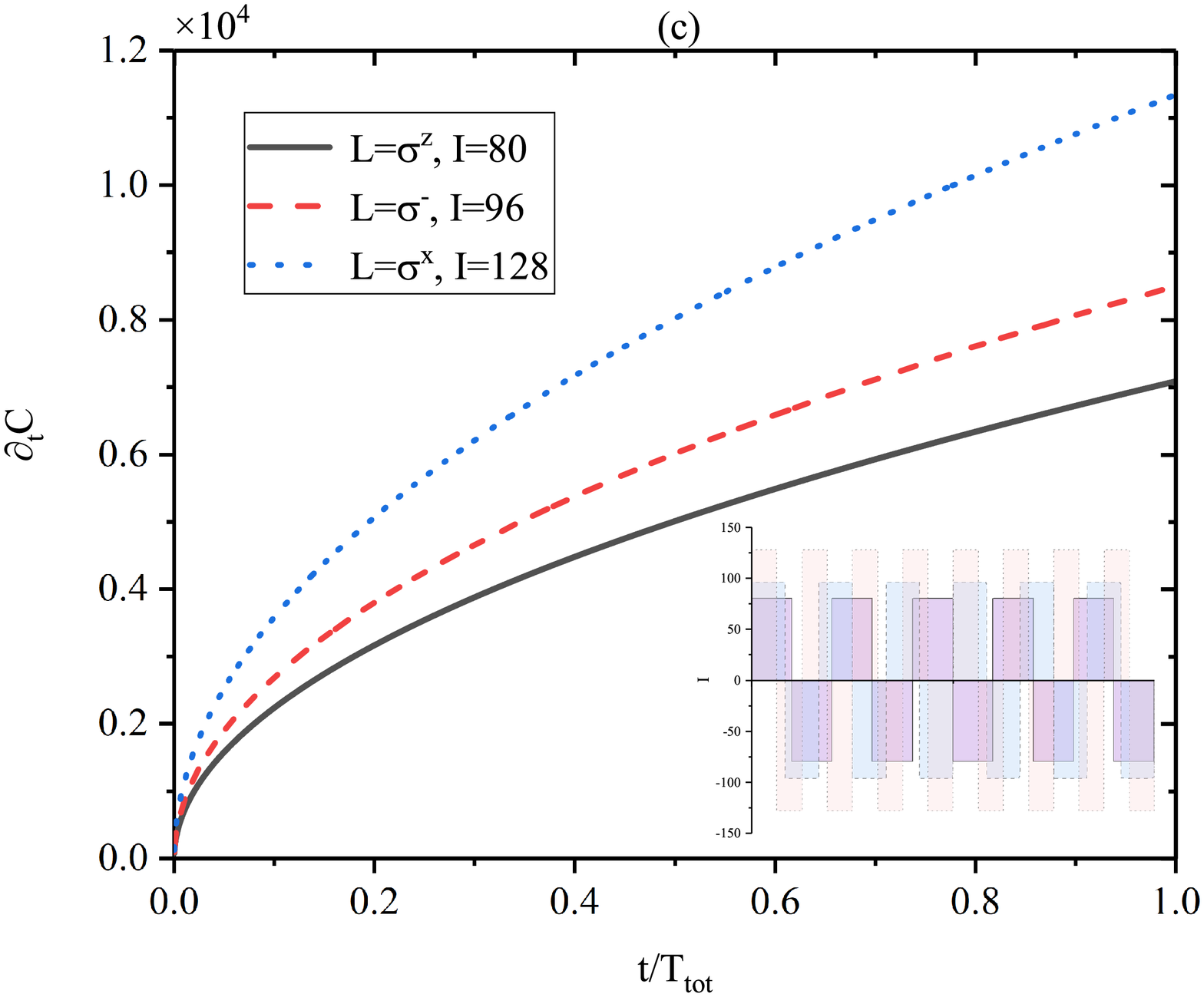}}
	\caption{(Color on line) (a) Controlled AEST in $L=\sigma^{x}$ open system: the fidelity $F$ as a function of rescaled time $t/T_{tot}$ for different control pulse intensity. (b) The instantaneous cost $\partial_{t}C$ as a function of rescaled time $t/T_{tot}$ under different control pulses. (c) The instantaneous cost $\partial_{t}C$ as a function of rescaled time $t/T_{tot}$ for different Lindblad operator. $\gamma=1$, $T=30$, $\Gamma=0.03$, $N=6$.} 
	\label{fig:2}
	
\end{figure}

From Eq.~(\ref{eq:23}), we can see that the control cost only depends on the control function $c(t)$ itself and the sum of $\left|f_{1,n}(t)\right|^2$. Here $f_{1,n}(t)=\langle n|\psi_{1}\left(t\right)\rangle$ is the transition amplitude of an excitation from the initial state to the $|n\rangle$ state \cite{Bose2003}. We take $L=\sigma^{x}$ as an example in this part. We point out that the types of the Lindblad operators do not change the conclusion of the paper. The environmental parameters are taken as $\gamma=1$, $T=30$, $\Gamma=0.03$. We use the rectangular pulses defined in Eq.~(\ref{eq:21}) and the intensities in Fig.~\ref{fig:2} all satisfy the control condition $I \tau=2 \pi$. In Fig.~\ref{fig:2}(a), we plot the fidelity $F$ versus the rescaled time $t/T_{tot}$ for different control intensities ($I=32, 64, 112, 128$). Clearly, The final fidelity increases with increasing control intensity. When $I=32$, the final fidelity is $F=0.971$. And when $I=64$, $F=0.995$. Fig.~\ref{fig:2}(b) plots the corresponding instantaneous cost $\partial_{t}C$ as a function of the rescaled time $t/T_{tot}$ for the same pulse intensities as plotted in Fig.~\ref{fig:2}(a). Fig.~\ref{fig:2}(b) shows that a higher fidelity $F$ corresponds to a higher control cost. So there exists a trade-off between the fidelity and the control cost. When the intensity is stronger enough, the fidelity is almost one. For example, when $I=112$, $F=0.999$. In this case, increasing the intensity just improves the control cost and it does not boost the fidelity ($I=128$, $F=0.999$). Then the minimum control intensity with minimum cost can be found to obtain certain transmission fidelity when the environmental parameters are fixed. To compare the effects of different types of Lindblad operator on the control cost, in Fig.~\ref{fig:2}(c) we plot the the instantaneous cost $\partial_{t}C$ as a function of rescaled time $t/T_{tot}$ for $L=\sigma^z, L=\sigma^-$ and $L=\sigma^x$, respectively. The final fidelity is set as $F=0.999$. In the inset of Fig.~\ref{fig:2}(c) we plot the strength of the pulse $I$ as a function of rescaled time $t/T_{tot}$. Clearly different pulse intensities are required to obtain certain fidelity for different types of $L$. The maximum control cost corresponds to $L=\sigma^{x}$, $L=\sigma^{z}$ is the least, and $L=\sigma^{-}$ is in the middle.  

\begin{figure}
	\centerline{\includegraphics[width=1.0\columnwidth]{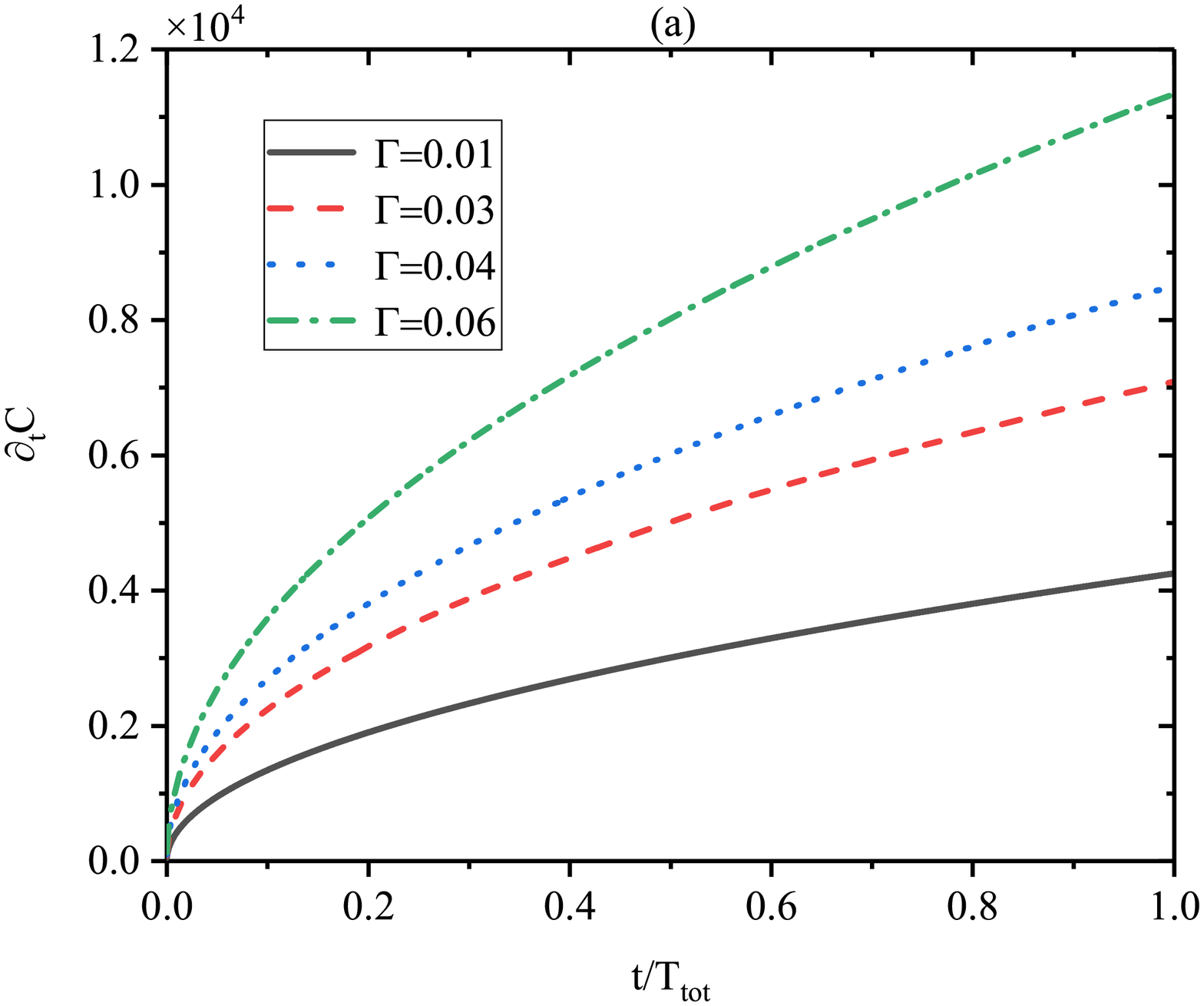}}
	\centerline{\includegraphics[width=1.0\columnwidth]{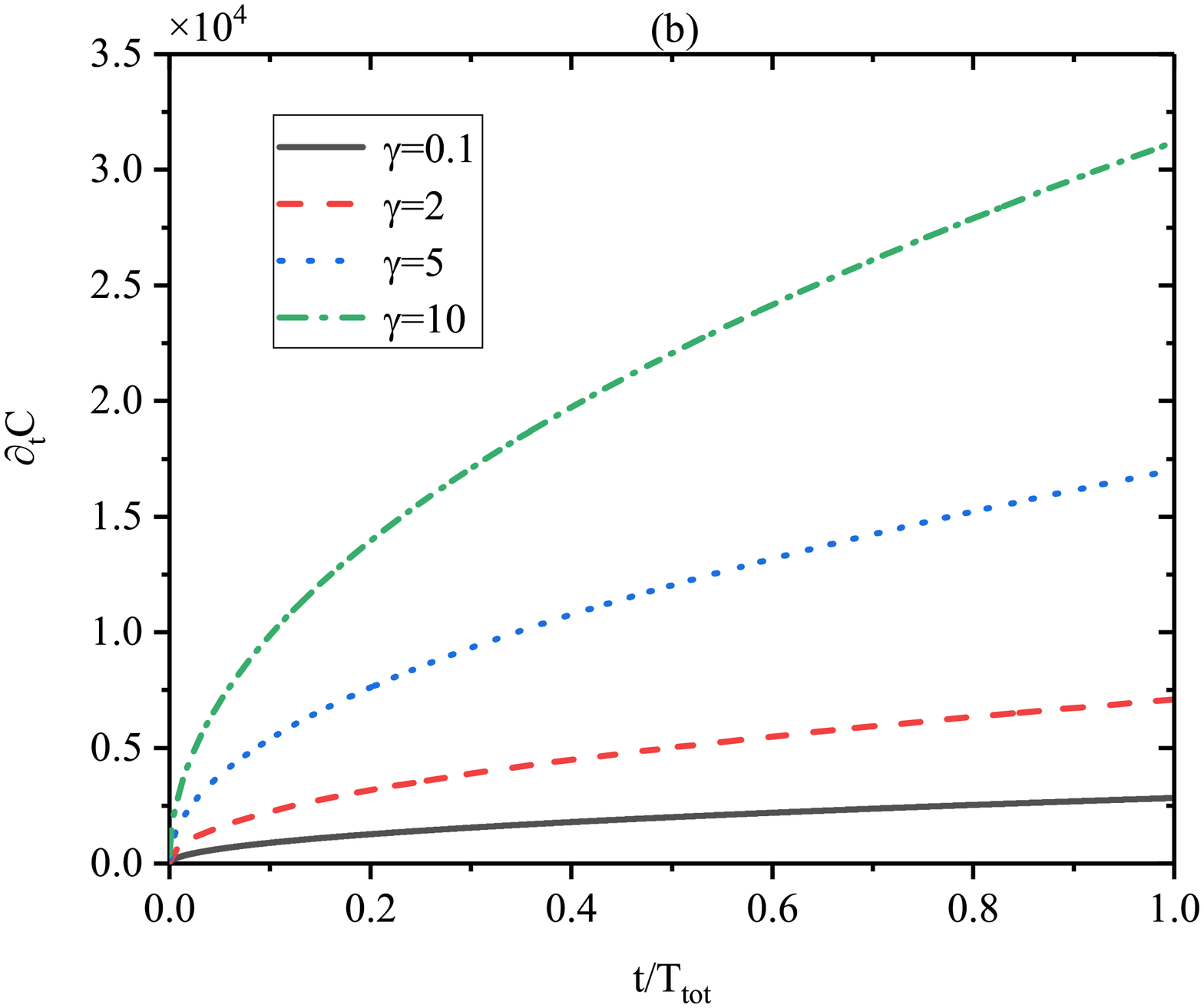}}
	\centerline{\includegraphics[width=1.0\columnwidth]{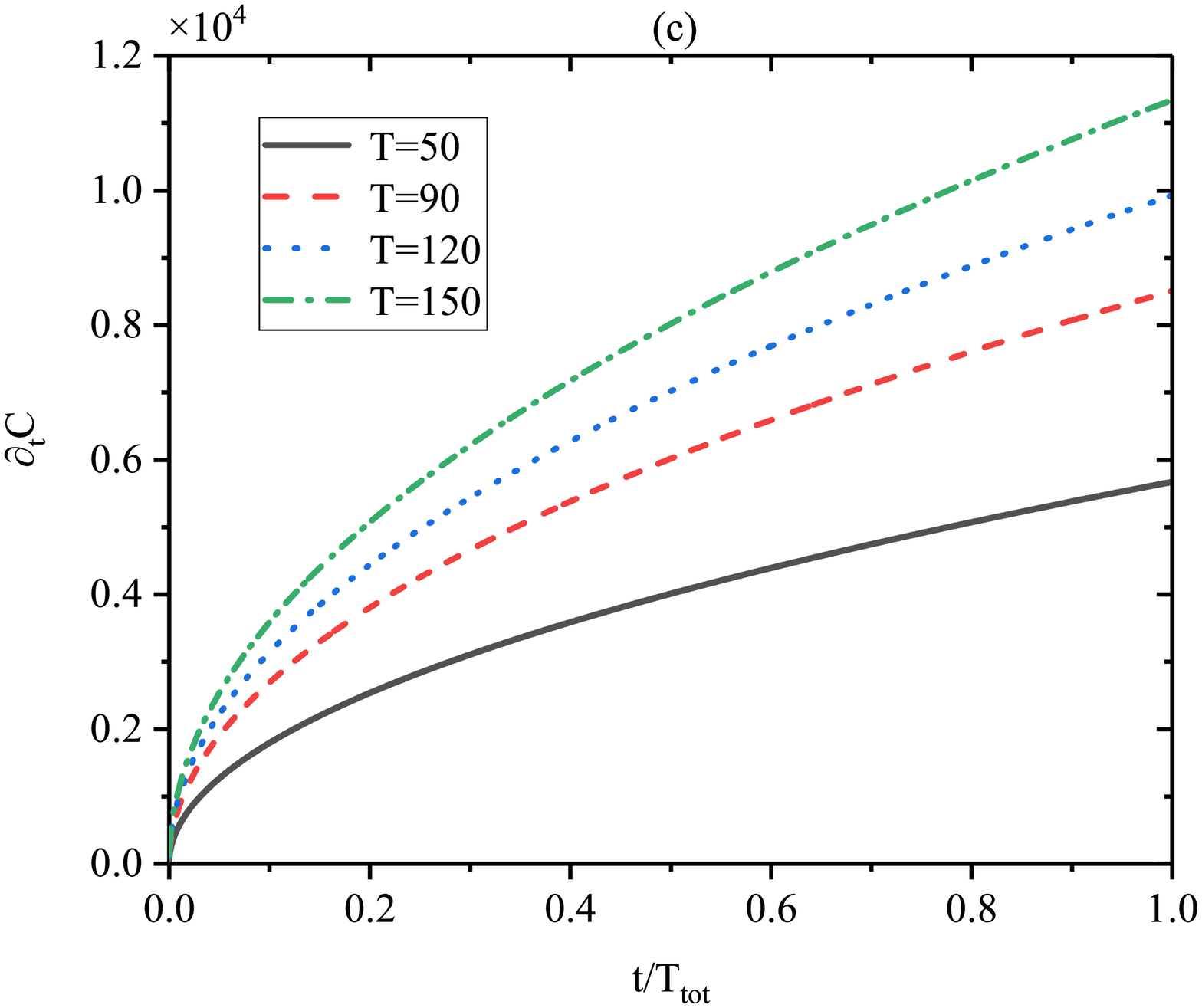}}
	\caption{(Color on line) The instantaneous cost $\partial_{t}C$ as a function of the rescaled time $t/T_{tot}$ for different parameters $\Gamma$, $\gamma$ and $T$, respectively. (a) $\Gamma$, $\gamma=2$, $T=30$; (b) $\gamma$ , $\Gamma=0.02$, $T=40$; (c) $T$, $\Gamma=0.03$, $\gamma=1$. The length of the chain is chosen to be $N=6$.}
	\label{fig:3}
\end{figure}

Next we consider the control cost when transferring a quantum state with certain fidelity. In Fig.~\ref{fig:3}(a)-(c) we plot the instantaneous cost $\partial_{t}C$ as a function of the rescaled time $t/T_{tot}$ for different parameters $\Gamma$, $\gamma$ and $T$, respectively. The final fidelity is $0.996$ under rectangular pulse control. As expected we observe that the control cost increases with increasing coupling strength $\Gamma$, the parameter $\gamma$, and bath temperature $T$ for a fixed final fidelity. Strong system-bath coupling and high temperature will destroy the quantumness of the system and as a result more control cost is required. Non-Markovianity from baths can be helpful to reduce the control cost notably from Fig.~\ref{fig:3}(b). Possible explanation maybe a memory environment has the property that the reverse information flow from baths to the system, with less loss of quantumness of the system.

\section{QSLT in controlled AEST}

In quantum physics, the uncertain relationship between the Heisenberg energy and the time limits the quantum speed of performing a quantum operation or QST \cite{Deffner2013a}. Recently the QSLT in open systems has been extensively studied \cite{Cimmarusti2015,Xu2014,Xu2014a,Mukherjee2013,Zhang2014,Taddei2013}.
In Ref.~\cite{Deffner2013}, Deffner and Lutz derived a unified bound for the QSLT in an open system by using the Cauchy-Schwarz inequality. Their approach defines a geometric Bures angle, $\mathcal{L}(\text{\ensuremath{\rho_{0},\rho_{t}}})=\mathrm{arccos}\left(\sqrt{\langle\psi_{0}|\rho_{t}|\psi_{0}\rangle}\right)$, which implies the ``distance" between the initial and final states. Then by using the time derivative of the Bures angle and $x\leq\left| x\right| $ \cite{Mirkin2016}, we can arrive an inequality
\begin{equation}
2\mathrm{cos}(\mathcal{L})\mathrm{sin}(\mathcal{L})\dot{\mathcal{L}}\leq\left|\langle\psi_{0}|\dot{\rho_{t}}|\psi_{0}\rangle\right|.\label{eq:24}
\end{equation} 
Using Eq.~\eqref{eq:24} and the Cauchy-Schwarz inequality we obtain
\begin{equation}
2\mathrm{cos}(\mathcal{L})\mathrm{sin}(\mathcal{L})\dot{\mathcal{L}}\leq\left\Vert \dot{\rho}\right\Vert.\label{eq:25}
\end{equation}  
Integrating Eq.~\eqref{eq:25} over time leads to the Mandelstam-Tamm type bound on the rate of quantum evolution, i.e., the QSLT $\tau_{QSL}$ 
\begin{equation} 
t\geq\tau_{QSL}\equiv\frac{1}{\Lambda_{t}}\mathrm{sin}^{2}[\mathcal{L}(\text{\ensuremath{\rho_{0},\rho_{t}}})].\label{eq:26}
\end{equation}
Here $\Lambda_{t}=\frac{1}{t}\int_{0}^{t}\mathrm{d}t$$\left\Vert \dot{\rho}\right\Vert$ represents the average of $\left\Vert \dot{\rho}\right\Vert$ over the actual driving time duration $t$. There are two viewpoints about QSLT. One focuses on the actual evolution time $t$, and the fastest evolution occurs when the actual driving time $t$ equals the QSLT $\tau_{QSL}$. It has been used in the discussion of entanglement assisted speedup of quantum evolution \cite{Batle,Frowis}. Another viewpoint is that the QSLT can be studied by first fixing the actual driving time $t$, which as been utilized to explore the non-Markovinity of the baths on the speed of quantum evolution \cite{Deffner2013,Zhu2015}. 
\begin{figure}
	\centerline{\includegraphics[width=1.0\columnwidth]{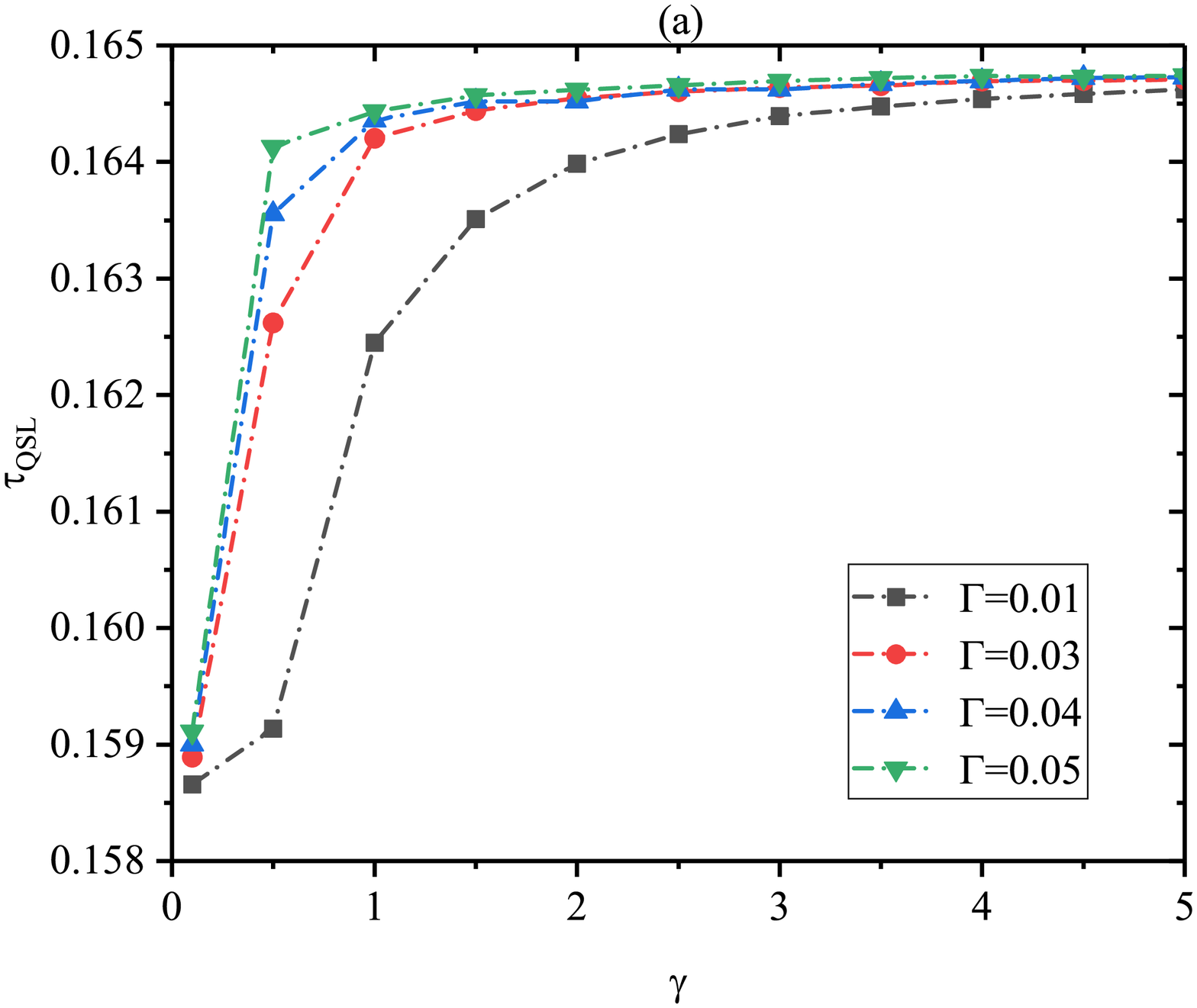}}
	\centerline{\includegraphics[width=1.0\columnwidth]{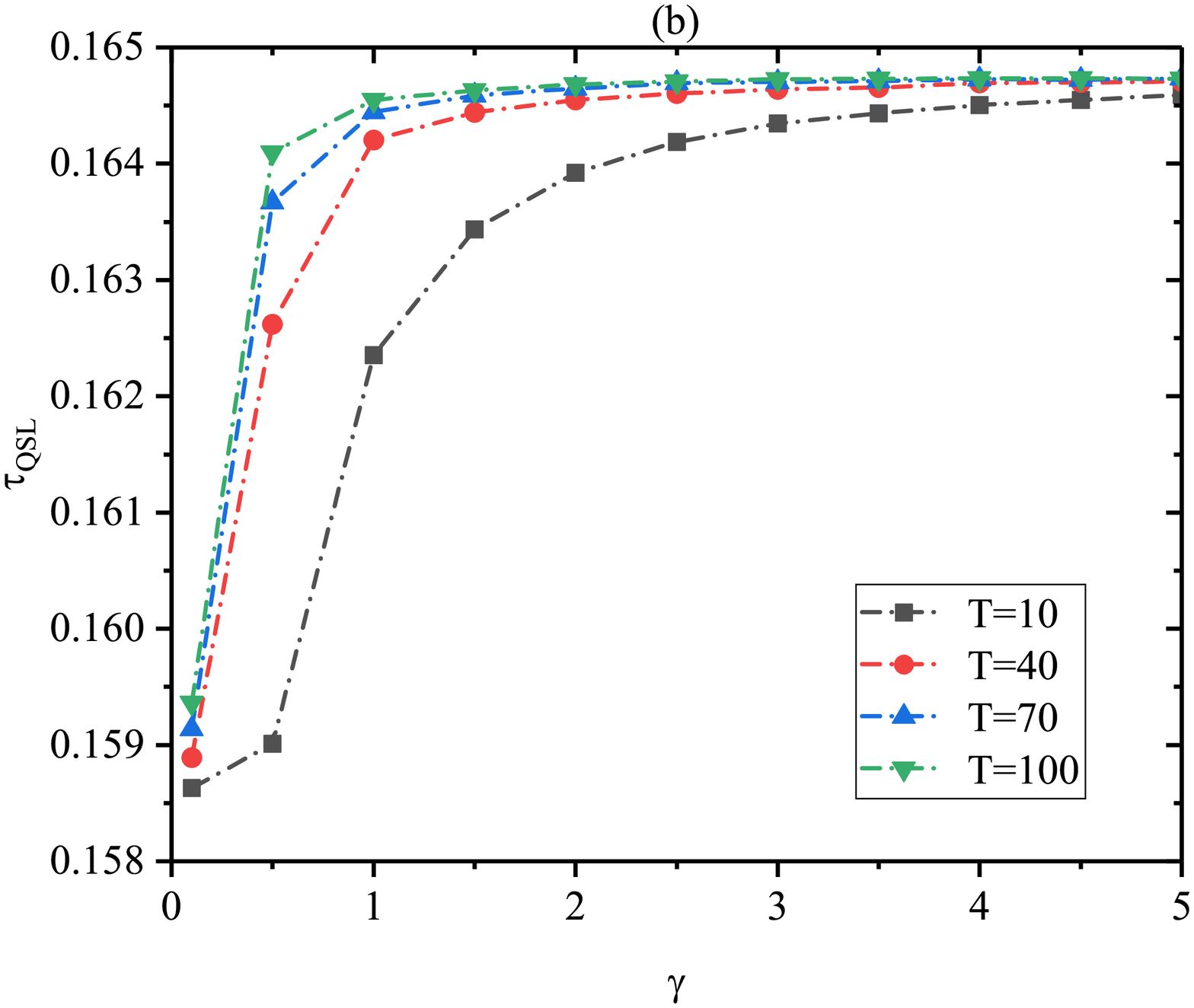}}
	\caption{(Color on line) The quantum speed limit time $\tau_{QSL}$ for various values of environment parameters. (a) $T=40$; (b) $\Gamma=0.03$. The length of the chain is chosen to be $N=5$.}
	\label{fig:4}
\end{figure}

Next we will discuss the influence of the environmental noise on the QSLT during the process of AEST. In our paper, we fix the actual driving time $t=\pi/4$ in Eq.~(\ref{eq:26}) since at this time AEST can be realized as mentioned above. Additionally, the final fidelity is taken as $F(t=\pi/4)=0.999$ for different parameters to guarantee that the initial state and the target state can be approximately the same. The values of $\tau_{QSL}$ for different environmental parameters are presented in Fig.~\ref{fig:4}. Note that the parameter windows are taken as $\Gamma \in [0.01,0.05]$, $T \in [10, 100]$, and $\gamma \in [0.1, 5]$ in Fig.~\ref{fig:4} due to the weak coupling approximations and the fidelity cannot reach 0.999 for baths with strong Markovianity. Fig.~\ref{fig:4}(a) and (b) plot $\tau_{QSL}$ versus $\gamma$ for different $\Gamma$ ($T=40$) and $T$ ($\Gamma=0.03$), respectively. They show that from the initial state $|1\cdots00\rangle$ to the target state $|0\cdots01\rangle$, $\tau_{QSL}$ increases with increasing $\Gamma$ or $T$, which means that weak system-bath couplings and low temperature will reduce the QSLT. Fig.~\ref{fig:4}(a) and (b) also show that non-Markovinity from the baths can help to shorten the QSLT and thus increase the capacity for potential speedup. The memory effect of environment is an important element for further reduction in QSLT \cite{Zhu2015}. 
	
\section{CONCLUSIONS}

Realization of AEST is a key to successfully complete quantum information processing tasks, however the environment noise around the communication channel will always destroy the state transmission fidelity. In this paper, we have explored the quantum state transmission through a spin chain under pulse control, where each spin of the chain is immersed in a non-Markovian, finite temperature heat bath. Using QSD equation approach, we numerically calculate the dynamics of this open system. We find that AEST can be obtained in non-Markovian and low temperature heat baths with weak system-bath coupling under a suitable pulse intensity and duration. We then consider the control cost and QSLT during the process of this AEST. We find that the cost and QSLT increases with increasing parameters $\Gamma$, $\gamma$ and $T$ for certain fidelity. Higher transmission fidelity corresponds to higher cost. The minimum cost has been found to obtain a certain target state. Moreover, it is also found that strong non-Markovianity (long memory time) can help to reduce the control cost and shorten the QSLT. Our findings will have potential applications for the control cost and information transfer in quantum device design.

\begin{acknowledgements}

We would like to thank E. Ya. Sherman for helpful discussions. This paper is based upon work supported by National Natural Science Foundation of China (Grant No. 11475160) and the Natural Science Foundation of Shandong Province (Grant No. ZR2014AM023).

\end{acknowledgements}
\nolinenumbers

\end{document}